\documentclass[conference]{IEEEtran}
\IEEEoverridecommandlockouts
\usepackage[english]{babel}
\usepackage[normalem]{ulem}
\newcommand{\href}[1]{#1} % does nothing, but defines the command so the print-optimized version will ignore \href tags (redefined by hyperref pkg).
\usepackage{notoccite}
\usepackage{ifthen}
\usepackage{graphicx,comment}
\usepackage[colorlinks=true, allcolors=blue]{hyperref}
\usepackage{soul}
\usepackage{amsmath}
\usepackage{amssymb}
\usepackage{algpseudocode}
\usepackage{enumitem}
\usepackage{mathrsfs}
\usepackage[table]{xcolor}
\usepackage{caption}
\usepackage{subcaption}

\usepackage{bbm}
\usepackage{algorithm}
\usepackage{tabularx}

\newboolean{PrintVersion}
\setboolean{PrintVersion}{false}
\usepackage{amsmath,amssymb,amstext} % Lots of math symbols and environments
\def\BibTeX{{\rm B\kern-.05em{\sc i\kern-.025em b}\kern-.08em
    T\kern-.1667em\lower.7ex\hbox{E}\kern-.125emX}}

\begin{document}

\title{Online Resource Management for the Uplink of Wideband Hybrid Beamforming System}

\author{\IEEEauthorblockN{Yuan Quan, Haseen Rahman, and Catherine Rosenberg,~\IEEEmembership{Fellow,~IEEE}}
\IEEEauthorblockA{\textit{Department of Electrical and Computer Engineering
} \\
\textit{University of Waterloo}, Canada\\
\{yuan.quan, hrcheriy, cath\}@uwaterloo.ca
}

}

\maketitle

\begin{abstract} 
This paper studies the radio resource management (RRM) for the \emph{uplink} (UL) of a cellular system with codebook-based \emph{hybrid beamforming}. We consider the often neglected but highly practical multi-channel case with fewer radio frequency chains in the base station than user equipment (UEs) in the cell, assuming one RF chain per UE. As for any UL RRM, a per-time slot solution is needed as the allocation of power to subchannels by a UE can only be done once it knows which subchannels it has been allocated. The RRM in this system comprises beam selection, user selection and power allocation, three steps that are intricately coupled and we will show that the order in which they are performed does impact performance and so does the amount of coupling that we take into account. Specifically,  we propose 4 online sequential solutions with different orders in which the steps are called and of different complexities, i.e., different levels of coupling between the steps. Our extensive numerical campaign for a mmWave system shows how a well-designed heuristic that takes some level of couplings between the steps can make the performance exceedingly better than a benchmark. 
\end{abstract}


\section{Introduction}
% \red{Hybrid beamforming (HBF) is the state of the art technology used in 5G communication systems. While fully digital beamforming has the potential to significantly improve the performance of communication systems, a commonly overlooked constraint for its implementation in multi-antenna configurations is the requirement of a costly and power-intensive radio frequency (RF) chain for each antenna.  
% Millimeter wave (mmWave) bands (30-300 GHz) have demonstrated significant potential to enhance the capability of the fifth-generation and beyond systems. However, mmWave bands face the challenges of more severe path loss, shadowing and penetration loss compared to microwave (sub-6GHz) bands  \cite{BF1}. Consequently, 

Massive multi-input multi-output (MIMO) and beamforming (BF) technologies significantly enhance spectral and energy efficiency for 5G and beyond systems. Fully digital BF (FDBF) exploits the full potential of beamforming \cite{MIMO1}, but the deployment costs and power consumption are high due to the need for one RF chain per antenna. As an alternative, hybrid BF (HBF) uses far fewer RF chains than antennas by incorporating analog BF (ABF) through phase shifters, enabling support for multiple data streams on the same resource, and achieving substantial BF gain \cite{hybrid-survey}. However, ABF also introduces practical constraints that must be addressed,  making radio resource management (RRM) in HBF systems fundamentally different from that in FDBF systems.

Most of the work on HBF systems has been on the downlink. The uplink (UL) has received very little attention in spite of the fact that it is more and more recognized as a bottleneck in terms of perceived quality of experience \cite{strinati}.

We consider the UL of an OFDMA-based cellular system that uses a codebook-based HBF, i.e., the base station (BS) and the UE codebooks are given and each user is mapped to a pair of preferred beams (one at the UE and one at the BS) during beam alignment.  %where the ABF vectors are selected from pre-designed codebooks is considered \cite{DL}. With fixed ABF vectors, the system can further save channel estimation (CE) overheads compared to architectures relying on full channel state information (CSI) \cite{UL2_LCHB,UL3_RLCH}. 
We assume a fully connected HBF architecture with $K$ RF chains at the BS and one RF chain at each UE \cite{hybrid-survey}.  %Each RF chain is used independently by a stream (user) in a subchannel. On the other hand, digital BF (a.k.a. precoding) involves the interaction between streams, which is left for future research.

In OFDMA, the bandwidth and time resources are organized into physical resource blocks (PRBs), each one consisting of one subchannel (the bandwidth is divided in $C$ subchannels) in a time slot. A frame is made of $N_T$ time slots, some are reserved for the downlink (DL) and some for the UL.  Each UE can send at most one stream per PRB with a maximum of $K$ UEs allowed to transmit in total in a PRB.  In the UL, the power available to each UE within a time slot is to be used in the PRBs (subchannels) allocated to that UE. This causes a coupling between user selection (USel)  and power allocation (PA) that does not exist on the DL. Thus, the solutions obtained for the DL in \cite{DL} are not directly applicable to the UL.  An important constraint to be taken into account is that when there is a need to select a set of beams of cardinality smaller than the number of preferred beams (recall that there is one preferred beam per user), then this beam set needs to remain the same for the entire time slot (i.e., the same set needs to be used for all PRBs in a time slot). We might not be able to select all preferred beams in all time slots because we allow $K<U$ where $U$ is the number of UEs in the cell. In that case,  per-time slot beam selection (BSel) is necessary since no more than $K$ beams can be selected which might be smaller than the total number of beams preferred by the users (typically the number of beams at the BS is much larger than $K$).  In this case, a UE can only be selected in a time slot if its preferred beam is selected in that time slot. Hence, per-channel solutions (e.g. \cite{UL1_LCMR,UL7_JUAH}) cannot be applied in this scenario. Of course, USel also needs to decide which subchannels to allocate to users who share the same preferred beams. Given a beam set, which users are selected in each beam in a given PRB affects the (inter-beam) interference at the BS and we will show that this cannot be ignored when performing USel. %This is coupled with user selection because the power is only allocated to the subchannels in which the UE is selected.} 

A comprehensive study of the UL RRM of a codebook-based HBF system considering all the above steps has not been done and forms the basis of this paper. Our two research goals are: 1) to understand the impact of the inter-beam interference and of the coupling between BSel, USel,  and PA on performance. 2) to propose an online efficient RRM solution for the system under consideration.
Our main contributions are summarized below.

% \sout{None of previous downlink and uplink solutions can be directly extended to our scenario because of the coupling between power allocation, user selection and beam selection. Hence, we propose RRM solutions that work in this case.}

\begin{itemize}[leftmargin=*]

  % \item \red{We formulate a joint optimization problem, which includes the mentioned RRM steps, on a time-slot basis. This is mainly to understand the complexity of RRM and highlight the couplings. This problem is highly nonconvex due to inter-beam interference and the couplings between ABS, US and PA.} %user selection and power allocation. 
  
  \item Because of the inter-beam interference and the coupling between BSel, USel and PA, optimizing the RRM jointly is difficult. Instead, we propose sequential online heuristics of low complexity that differ in the steps being used and in the sequence order of the different RRM steps, the order of which significantly impacts the performance.
  \begin{itemize}
      \item We first propose a \emph{benchmark} that does not consider the coupling between RRM steps, ignores inter-beam interference, performs BSel in round-robin and uses the state-of-the-art per-beam user selection (i.e., a per-beam proportional fair uplink scheduler) as well as equal power allocation (EPA).
      \item Our first heuristic, $S_0$, differs from the benchmark only in one step, i.e., it uses an enhanced USel algorithm, since it turns out that the state-of-the-art USel does not perform well for a small number of UEs. $S_0$ shows significant improvement over the benchmark. 
      \item We then further improve the performance in $S_1$ by making BSel \emph{load-aware}, i.e, by performing USel per beam, first, as in $S_0$ and selecting the maximum allowable beams that yield the highest individual performance metric. We keep the same USel and PA as in $S_0$.
      \item Finally, in our ultimate heuristic, $S_2$,  we add a final step to $S_1$ to take inter-beam interference into account by revisiting USel once BSel is done.
  \end{itemize}
  \item Our extensive numerical study on a mmWave band shows that $S_2$ is performing up to 6 times better than $S_1$ which is performing up to 91\% better than $S_0$ which is performing up to 30\% better than the benchmark (70\% gain at least from $S_0$ to $S_2$), showing the importance of taking the couplings and the inter-beam interference into account as well as having a scheduler that performs well for a small number of users (since typically the number of users per beam will be small).
  %: per-beam user selection, load-aware (LA) beam selection, interference-aware (IA) user dropping and EPA. We run user selection before beam selection to make it LA instead of independent and load-unaware (LU) beam and user selections (e.g., round-robin). %To analyze the impact of each block, we compare the heuristic’s performance with other block combinations, enabling a step-by-step understanding of each block’s contribution to overall system performance.
  
  %\item The user selection selects UEs independently for all preferred beams. Then we select beams (as many as we can) that realize the highest weighted sum rates based on per-beam results. Hence the beam selection is LA by considering loads (weighted sum rates) of beams. Numerical results indicate the necessity of LA beam and user selections.

  %\item Since per-beam user selection ignores inter-beam interference, we design an IA user deselection block to drop UEs potentially causing significant interference to other UEs at the BS after beam selection. Results show that this block is necessary since it largely improves the performance when multiple UEs share a subchannel.
  
  \item Finally, we compare the performance of $S_2$  when we add a final step where we perform power allocation based on water-filling (WF) and show that this step is not improving the performance.
  
\end{itemize}

The remainder of the paper is organized as follows. Section~\ref{Sysmodel} describes the system model.  Section~\ref{RRM} provides our RRM solutions. Section~\ref{Results} presents our simulation results. Section~\ref{Conclusion} concludes the paper.

\section{System Model} \label{Sysmodel}
We study the UL RRM of a single-cell massive MIMO system. The system uses a fully connected HBF architecture. Specifically, there are $N_b$ antennas, $K$ RF chains and $N_b \times K$ phase shifters at the BS, and $N_u$ antennas, one RF chain and $N_u$ phase shifters at each UE. Each RF chain is connected to all $N_b$ (resp. $N_u$) antennas by $N_b$ (resp. $N_u$) phase shifters at the BS (resp. UE). We assume $K\ll N_b$ and allow $U>K$ where $U$ is the number of UEs. Assuming the system is operated in a time division duplex (TDD) mode where the full bandwidth is used for either UL or DL in a time slot.
\subsection{The Inputs to RRM}
In addition to fixed system parameters, all the online solutions rely on three input sets that are updated every time slot or less as will be discussed next. The first set contains a pair of preferred beams for each UE (one at the BS and one at the UE). These pairs are fixed for a certain number of time slots depending on how often beam alignment (BA) is performed (see Section~\ref{BA}). %can remain fixed for a specified number (this is to take into account different practical cases, e.g., one or $N_T$) of time slots and are chosen through beam alignment (BA). 
The second input set contains the effective channel state information (ECSI) vector ($U \times 1$) per PRB for each UE. These vectors are obtained by channel estimation (CE) using the UE's preferred beams (see Section~\ref{subsec:ECSI}). Full CSI is not needed. Lastly,  RRM steps uses per-UE weights derived from the average rate obtained in past time slots to account for proportional fairness.  This will be introduced in Section~\ref{FN}. Next, we introduce enough details on the BA and CE procedures to understand how the inputs are obtained.

\subsection{Beam Alignment}\label{BA}

We assume that ABF codebooks $\mathcal{C}_b$ at the BS and $\mathcal{C}_u$ at the UE have been selected in a planning phase, where $\mathcal{C}_b=\{\mathbf{w}_j \in \mathbb{C}^{N_b \times 1}:\lVert \mathbf{w}_j \lVert ^2 =1,j=1,...,B_b\}$ and $\mathcal{C}_u=\{\mathbf{v}_j(u) \in \mathbb{C}^{N_u \times 1}:\lVert \mathbf{v}_j(u) \lVert ^2 =1,j=1,...,B_u\}$ with $B_b$ (resp. $B_u$) is the size of the codebook at the BS (resp. UE). Under the assumption that all UEs use the same codebook, we remove index $u$ in the following. BA matches each UE with the best pair of beams, one from $\mathcal{C}_b$ and one from $\mathcal{C}_u$. We call this pair,  the UE's preferred pair of beams. BA can be realized by various techniques (e.g., \cite{DL}, \cite{BA3}). In this paper, we select the beam pair $(\mathbf{w}_u^*,\mathbf{v}_u^*)$ that provides the highest average channel gain for UE $u$ \cite{DL}. We define $\mathcal{B}_p=\{b_1,...,b_{|\mathcal{B}_p|}\}$ as the set of indices of all preferred beams at the BS. We also have the mappings $b(u)$ from any UE $u$ to the index of its preferred BS beam and $u(b)$ from any BS beam $b$ to the UE(s) preferring BS beam $b$. Typically BA is done once every $N_T$ time slots where $N_T$ is relatively large. Our heuristics do not make any assumption on the value of $N_T$. %T\st{We assume the BA results remain the same for $N_T$ time slots. The RRM heuristics that we propose work for any value from one to $N_T$.} \hl{This is something for the numerical section}

\subsection{Effective Channel Estimation} \label{subsec:ECSI}

Following BA and before RRM, each UE $u$ transmits pilot signals using ABF vector $\mathbf{v}_u^*$,  $1\leq Q \leq C$ times per time slot (i.e., for each block of $C/Q$ PRBs in a time slot). The BS measures the pilot signals of each UE and compute the effective channel coefficient in any PRB $c$ of block $q$, as  $g_{c,n,u}^\text{eff}=\left({\mathbf{v}^*_u}\right)^H \mathbf{G}_{q,u}\left(\mathbf{w}_n^*\right)$ where $\mathbf{G}_{q,u}$ is the channel matrix between the BS and UE $u$ in block $q$. If $n=u$, it denotes the effective signal channel between UE $u$ and the BS. If $n\not=u$, it denotes the effective interference channel between UE $n$ and the BS due to UE $u$ transmission to the BS. Our heuristics do no make any assumption on the value of $Q$. 

Note that the RRM techniques that we are about to introduce are independent of the BA and CE methods. In this paper, we posit the perfection of CE.%, i.e., they are the channel coefficients after applying the ABF vectors. 

% The rest of the derivation of effective channels depends on the existence of DBF. Therefore, we consider two cases. \com{[[The notation of $h$ is not good. What we call effective channel does not have a superscript ``eff'' and what we do not call effective channel does have that superscript. It is weird and will certainly confuse the reader.]]}

\subsection{Fairness} \label{FN}
The RRM objective is to provide proportional fairness (PF) over a certain horizon. Recall that all our solutions works on a time slot basis. Specifically, in line with what is being done in \cite{DL,PF1}, at the beginning of time slot $i$, let $R_u(i)$ be the average throughput received by $u$ over the past $W$ time slots. Then, the objective in that time slot is to
maximize the product of the UEs' rates over the past (and the present) $W+1$ time slots \cite{DL}, i.e., 
%The objective function that we consider at the beginning of the $t$-th CTI is the proportional fairness one in \cite{PF} which maximizes of the geometric mean of the throughput received by each UE over $W+1$ CTIs, i.e.,
\begin{align}\label{OF}
    & \max \prod_{u \in \mathcal{U}} \left( WR_u(i)+\lambda_u(i) \right) \approx \max \sum_{u\in \mathcal{U}} \frac{\lambda_u(i)}{W R_u(i)},
\end{align}
where $\lambda_u(i)$ is the rate of UE $u$ in time slot $i$ which is the sum of the rates seen in each PRB of that time slot. This can be approximated by a weighted sum rate maximization where the weight of user $u$ is $w_u(i)=1/R_u(i)$ as long as $WR_u(i) \gg \lambda_u(i)$, which is reasonable to assume. At the end of time slot $i$, we use a moving average of size $W$ to update $R_u(i)$ as $R_u(i+1)=(1-1/W) R_u(i)+\lambda_u(i)/W, \forall u \in \mathcal{U}$ and the weight becomes $w_u(i+1)=1/R_u(i+1)$.  For brevity of notation, we will remove the time slot index $i$. %We further approximate the objective function by $\sum_{u\in \mathcal{U}} \frac{\lambda_u}{W R_u}$ assuming $WR_u \gg \lambda_u$ and using the approximation $\log(1+x) \approx x$ for $x \ll 1$ \cite{DL}.

Finally, note that the natural performance metric for PF is the geometric mean of the rates $GM$, as discussed in \cite{DL}, i.e., given a time horizon of $N_T$ time slots,
\begin{align}
    GM = \left(\prod_{u=1}^U \frac{1}{N_T} \sum_{i=1}^{N_T}  \lambda_u(i)\right)^\frac{1}{U}.
\end{align}
\subsection{Rate Function} \label{subsec:ratefunc}
Given a selected beam set and given  that user set $z$ has been selected in a given PRB $c$, 
the signal-to-interference-plus-noise ratio (SINR) of UE $u \in z$  is
\begin{align}
    & \gamma^{c,u}(z)=\frac{|g_{c,u,u}^\text{eff}(z)|^2 p^{c,u}(z)}{
    \sum\limits_{\begin{subarray}{c} n\in z\\ n \neq u \end{subarray}} {|g_{c,u,n}^\text{eff}(z)|^2 p^{c,n}(z)} + \sigma_\text{PRB}^2}, \label{C7}
    %&\forall u \in \mathcal{U},  \forall \ell \in \mathcal{L}, \forall q \in \mathcal{Q}, \forall z_\ell \in \mathcal{M}_\ell \nonumber
\end{align}
where $\sigma_\text{PRB}^2$ is the noise power on a PRB and $p^{c,u}(z)$ is the power allocated to PRB $c$ by UE $u$. Hence, to obtain the SINRs, we need to do power allocation, i.e., distribute the UE's power budget to its allocated PRBs.

%within a \hl{block [we do not use the term RBL]} $q$

We use a rate function based on practical modulation and coding schemes (MCSs) rather than Shannon capacity formula to map the SINR of a UE in a given PRB to the maximum achievable data rate in bps in this PRB. Specifically, we model the rate function as a piecewise constant function of SINR, consisting of $L$ levels corresponding to the $L$ MCSs, i.e.,
\begin{equation} f(\gamma)=B_c\big(s_1\mathbbm{1}_{[\Gamma_1,\Gamma_2)}(\gamma) + %e_2 \mathbbm{1}_{[\Gamma_2,\Gamma_3)}(\gamma)+
  \cdots + s_{L} \mathbbm{1}_{[\Gamma_{L},\infty)}(\gamma)\big), \label{eq:true_rate}
\end{equation} 
where $B_c$ is the bandwidth of a PRB in Hz, $\mathbbm{1}_A(x)$ is the indicator function, which equals 1 if $x \in A$ and zero otherwise, $\Gamma_l$ denotes the SINR decoding threshold for MCS $l$ and $s_l$ is the spectral efficiency of MCS $l$ measured in bps/Hz. The set of MCSs is adopted from 3GPP TS 38.214 \cite{3GPP}.  %\red{In the caption you introduce f hat but you do not introduce it in the text. Not good.}
% \vspace{-5pt}

\section{Radio Resource Management} \label{RRM}
\begin{figure}
    \centering
    \includegraphics[width=80mm]{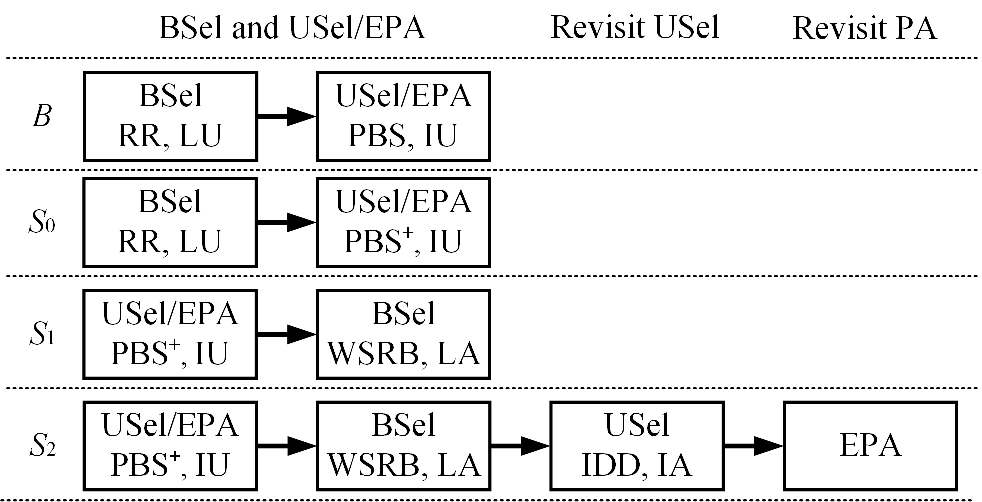}
    \caption{Our RRM solutions: RR: round robin, IU: interference unaware, IA: interference aware LA: load aware, LU: load unaware, EPA: equal power allocation, PBS: per beam selection, WSRB: weighted sum rate beam selection.} 
    \label{Solutions}
\end{figure}

In this section, we present our four online solutions (including the benchmark). Our goal is to identify what are the aspects of the various RRM steps that have a significant impact on the performance of the system. Recall that the UL RRM includes the following steps:
\begin{enumerate}[leftmargin=*]

\item \textit{Beam selection (BSel)}: Let \( L = \min(|\mathcal{B}_p|, K) \), representing the maximum number of beams we can select. In each time slot, up to \( L \) beams are selected for transmission. The selected beams are fixed for all subchannels in a time slot. Recall that beams may have different ``loads" (a term we use loosely) because the numbers of UEs preferring each beam are different.

\item \textit{User selection (USel)}: Within a time slot, for each beam of the selected subset of beams, at most one UE is to be selected in each subchannel. This is coupled with beam selection because a UE can only be selected if its preferred BS beam is activated in the time slot. Note that there might be significant inter-beam interference at the BS  since multiple  (up to $L$) UEs can transmit on the same subchannel.

\item \textit{Power allocation (PA)}: The power budget of each UE needs to be allocated to the subchannels on which it is selected for transmission. Hence, PA is clearly coupled with USel.
% \item \textit{Modulation and coding scheme (MCS) selection}: The adaptive MCS function used in practical systems decides the rate of transmissions for the selected UEs by mapping the signal-to-interference-plus-noise ratio (SINR) obtained to a pre-defined set of spectral efficiencies.
\end{enumerate}
%We first identify that the joint optimizing all the RRM steps is highly non-convex due to inter-beam interference and the coupling between BSel, USel and PA.
%\sout{ we first formulate the UL RRM optimization problem of the codebook-based HBF systems including beam selection, user selection and PA. The formulation will show that the problem is highly non-convex due to inter-beam interference and the coupling between user selection and PA. Rather than solving the problem directly.}
Next, we propose four sequential online heuristics (one of them being the benchmark) of low complexity that differ in the steps being used and in the sequence order of the different RRM steps, the order of which significantly impacts the performance.
%Next, we describe our four solutions the benchmark RRM solution which does not consider the coupling between RRM steps with load-unaware BSel and interference-unaware USel. Second, we improve the USel from the literature with a persistent search strategy. Third, we place USel before BSel and make BSel LA. Finally, we revisit the USel by dropping UEs causing high interference, which makes our USel IA. We will demonstrate the impact of each step by comparing the four heuristics. We use EPA or WF as the PA solution. Note that the RRM unit is a time slot. The inputs to RRM are $\mathcal{B}_p$, $b(u)$, $u(b)$, $R_u$ and $g_{q,n,u}^\text{eff}$ for all UEs, channels and preferred BS beams besides fixed system parameters and functions. 
Fig.~\ref{Solutions} shows the sequence of steps of all four solutions and the main attributes of the steps.

\subsection{RRM Benchmark}
We first build a basic RRM solution as the benchmark (called $B$ in the following). Given a time slot:
\begin{itemize}
    \item For BSel, we select a beam set of size $L$  from $\mathcal{B}_p$ in a round-robin (RR) fashion (one beam set per time slot). This BSel is load-unaware (LA), i.e., it does take into account the fact that some beams might be preferred more than others.
    \item For USel and equal power allocation (EPA) jointly: Given the selected beam set in a time slot, we select one UE per PRB  for each beam independently without considering the inter-beam interference, hence USel is interference-unaware (IU). Specifically, for each subchannel within a time slot, we select at most one UE from those preferring the given beam ($u(b)$), assuming that this beam is the only active one in that time slot. Focusing on one beam, this is similar to single-user (SU) MIMO user scheduling. Hence, the solution in \cite{PerBeam}, called per beam selection (PBS),  can be applied to generate the per-beam user selection results. PBS selects UEs in a greedy way taking their weights into account. In each step, it first assumes each UE is given one more subchannel among those not yet allocated (its highest gain subchannel), and calculates a revised weighted rate for each UE in the time slot assuming EPA. Under this assumption, PBS finds the UE whose weighted rate increases the most and gives its preferred subchannel. The algorithm stops if there are no more subchannels to give or if none of the UEs see an increase in rate. 
\end{itemize} 
Note that PBS gives us not only a user set per subchannel but also a power allocation (based on EPA). Hence, we have everything we need to compute the SINRs in this time slot, the rates and update the weights.

\subsection{$S_0$: Improving on PBS}
Extensive simulation campaigns have shown that PBS does not perform well when the number of users preferring a beam is small (which is often the case in practice, e.g, if you have a cell with 40 users and 32 beams, there are just a few users preferring a specific beam). In that case, we found out that PBS might stop allocating subchannels to users too early, and that even if adding a subchannel might not increase the rate seen by a user, adding more than one might do it. 
%However, the algorithm in \cite{PerBeam} is designed for SU-MIMO applications where the number of UEs is large and subchannels are mostly fully occupied by UEs. In our case, the number of UEs preferring a beam can be small (e.g., one). Hence the algorithm may stop early and there will be spare subchannels for a beam. 
We implement a persistent search strategy in which we do not stop allocating subchannels to a UE as soon as we see a decrease in its rate. This strategy can improve the performance by up to 27\% based on our results. Specifically, We propose an enhanced version of PBS, PBS$^+$, in which the search, for a UE, is stopped once we see  $X$ cumulative iterations with a decrease in rate. We describe PBS$^+$ in Alg.~\ref{PerBeam} for a given selected beam $b$ and a given time slot. PBS$^+$ can be run independently for all selected beams because a UE selects one preferred BS beam. Note that the inputs to the algorithm are the set $u(b)$ of users preferring beam $b$, $X$, the weights  $w_u$, per user power $P_{\text{UE}}$ and the set of all the effective channel gains $\mathcal{G}^b_{\text{eff}}=\{g_{c,n,u}^\text{eff}, \forall n,u \in u(b)$\} and for all subchannels $c$.

\addtolength{\topmargin}{0.05in}

\begin{algorithm}
\caption{\emph{PBS$^+$ on beam $b$}, Inputs: $u(b)$, $X$, $w_u$, $P_{\text{UE}}$, $\mathcal{G}^b_{\text{eff}}$}
\label{PerBeam}
\begin{algorithmic}[1]
%\For{each beam}
    \State Set the stop flags and the sum-rate SR(u) of all UEs in beam $b$ to zero.
    \While{there are any remaining subchannels to allocate in this time slot \textbf{and} UEs whose stop flag is no larger than $X$}
    \For{each UE in this beam whose stop flag is no larger than $X$}
        \State Select the best remaining channel ($\max_c g_{c,u,u}^\text{eff}$) for this UE. \label{lp1}
        \State Calculate SR(u) of this UE in this time slot assuming  EPA and that this channel is allocated to it.
        \State Increase its stop flag by one if the sum rate of this UE does not increase.
    \EndFor
    \State Select the UE whose weighted sum rate increases the most and allocate its best remaining channel to it.
    \State Update the sum-rate SR(u) of that UE. \label{lp2}
    \EndWhile
    % \State Record the UE (if any) allocated to each subchannel and the power allocated to this channel. \hl{Calculate and record the weighted sum rate of all UEs in the beam ignoring interference. [why?]} \label{rate}
%\EndFor
% \State Set the average throughput of all UEs back before step 1. \hl{[why?]}
 
\end{algorithmic}
\end{algorithm}

% \st{The output of Alg.1 is the selected UE (could be no UE) in each subchannel of a time slot for all preferred beams, and the (temporary) power allocated to the subchannels occupied by a UE. We can further accelerate Alg.1 by adding 
% $\frac{C}{Q}$ subchannels per loop (steps  to ). However, it's essential to attempt adding a single subchannel afterward, as a UE may not consistently have sufficient power to support the addition of 
% $\frac{C}{Q}$ subchannels at once.}

$S_0$ has the same sequence of steps as the benchmark. The only difference is that we use PBS$^+$ instead of PBS as the joint per beam USel and PA. %By replacing the USel in the Benchmark with Alg.~\ref{PerBeam}, we have the second solution $S_0$.

\subsection{$S_1$: Load-Aware Beam Selection}
In the previous two solutions, we select beams first in a RR fashion without considering the loads (i.e., the numbers of UEs preferring a beam) of a beam which affects the overall fairness. In the third solution $S_1$, we restructure the RRM sequence by moving USel before BSel. This approach allows for an estimation of load before performing BSel, enabling load-aware BSel. Note that if $|B_p| < K$, all preferred beams can be selected and there is no difference between $S_0$ and $S_1$. % rather than a load-unaware approach, as seen in $B$ and $S_0$.

Specifically, at the beginning of a time slot, we first run PBS$^+$ independently on each beam in $B_p$. Then we compute the per-beam weighted sum rate and select the 
$L$ beams with the highest weighted sum rates. This LA BSel method is referred to as weighted sum rate BSel (WSRB). %This BSel is load-aware, accounting for the “load” of each beam by estimating the potential achievable weighted sum rate. We will compare LA and LU manners in Section~\ref{Results}.

\subsection{$S_2$: Interference-Aware Two-Step USel}
In the previous three solutions, the per-beam user selection does not consider the inter-beam interference in a subchannel. Hence interference is ignored in the whole RRM process, making the approach interference-unaware (IU). Nevertheless, inter-beam interference can significantly affect the rate achieved by each UE when multiple UEs share the same subchannel. To address this, we propose an interference-aware USel approach that is done in two steps. Specifically, referring to $S1$, we add a step, called interference down dropping (IDD),  after BSel to revisit USel, i.e., to possibly deselect UEs in a subchannel if they generate large interference at the base station (e,g, a level $I_D$ times larger than the signal strength of any other UE within the subchannel).  

To explain IDD in more detail, we need some notations. Note that IDD is done independently per subchannel.
Given a subchannel $c$. let $I_{c,n,u}=\frac{|g_{c,n,u}^\text{eff}|^2 p^{c,u}}{|g_{c,n,n}^\text{eff}|^2 p^{c,n}}$ be the ratio between the interference impacting the transmission between UE $n$ and the BS  due to the transmission of UE $u$ to the BS and the signal power of UE $n$ at the BS. It is set as zero if $n=u$. We further define $\mathcal{I}_{c,u}=[I_{c,n_1,u},...I_{c,n,u},...,I_{c,n_{|z_c|},u}]^T$ for $\forall n \in z_c$. 

 The details of IDD are described in Alg.~\ref{IDD} for subchannel $c$. The inputs are the user set $z_c$ obtained after the first two steps in $S_1$, $I_D$ as well as $\mathcal{I}_{c,u} \forall u \in z_c$. 

\begin{algorithm}
\small
\caption{{\it Interference Down Dropping for subchannel $c$}, Inputs: $\{\mathcal{I}_{c,u}, \forall u \in z_c\}$, $I_D$ }\label{IDD}
\begin{algorithmic}[1]
    \For{each  $u \in z_c$}
        \State Verify if any element in $\mathcal{I}_{c,u}$ is larger than $I_D$.
        \State If yes, record the UE.
    \EndFor
    \State Deselect the recorded UE(s) in this subchannel.
\end{algorithmic}
\end{algorithm}

This algorithm deselects UEs potentially causing too much interference to other UEs at the BS. We refer to this process as IDD. Once, IDD has been applied to all subchannels, we redo EPA. In Section~\ref{Results}, we will compare the performance with and without this process to understand its significance.

\subsection{Ultimate water filling (WF)-based Power Allocation}

We can add an extra step at the end of $S_2$, in which we apply water filling (WF)-based PA \cite{WF} to allocate the power of each UE to the subchannels they occupy in a time slot. This step adds complexity but may be necessary to improve performance. We will compare the performance with and without this extra step in Section~\ref{Results} for $S_2$.

%In the end, we calculate the SINR of each UE in each subchannel and map it to rate according to the MCS-based rate function in Section \ref{subsec:ratefunc}. We will compare the performance of EPA and WF 

\section{Numerical Results} \label{Results}

In this section, we present the numerical results for our four solutions, namely the benchmark, $S_0$, $S_1$ and $S_2$ in a mmWave single cell. 
Specifically, we consider the uplink of a small cell with a radius of \( r = 75 \) m, with UEs uniformly distributed excluding a 6 m radius around the BS where path loss models are invalid. The BS height is \( 10 \) m. Each UE has a power budget of \( P_{UE} = 7 \) dBm and the noise power spectral density $N_0$ is -174 dBm/Hz. The BS and UEs use uniform linear antenna arrays (ULAs) with 128 and 16 antennas, respectively, spaced in half-wavelength. The system operates at \( 28 \) GHz with a \( 100 \) MHz bandwidth, divided into \( C = 132 \) subchannels of $B_c=720$ kHz. We assume \( Q = 22 \), i.e., a coherence bandwidth of 4.32 MHz which is typical in a mmWave small cell \cite{DL}.  We set the horizon to be \( N_T = 100 \) time slots. We generate ABF codebooks for the BS and UEs using the method in \cite{CODEBOOK1}.% \st{with M = 15 and T = 3} \hl{What are M and T?}. \blue{[Parameters in \cite{CODEBOOK1}, not necessary.]}

%\red{We assume that BA is only done once at the beginning of the $N_T$ time slots.} \blue{[We mentioned this later when defining realization.]}

We adopt a wideband channel model which incorporates multipath effects, i.e., scattering clusters and spatial paths in each cluster \cite{DL}. The MIMO channel between BS and UE $u$ at block $q$ is defined as the following $N_u \times N_b$ matrix:
\begin{align}
    \mathbf{G}_{q,u} = \frac{1}{\sqrt{N_\text{path}}} \sum_{d=1}^{N_\text{cluster}} \sum_{l=1}^{N_\text{path}} g_{d,l}^{u} e^{-j 2\pi \tau_{d,l}^u f_q} \mathbf{a}_\text{RX}(\phi_{d,l}^u) \mathbf{a}_\text{TX}^\text{H}(\theta_{d,l}^u),
\end{align}
where $N_\text{cluster}$ is the number of clusters, $N_\text{path}$ is the number of paths in each cluster, $g_{d,l}^{u}$ is the  coefficient of the path $l$ of cluster $d$, $f_q$ is the center frequency of block $q$, and $\tau_{d,l}^u={\tau_0}_{d}^u+{\tau_1}_{d,l}^u$ in which ${\tau_0}_{d}^u$ is the group delay of the $d$-th cluster and ${\tau_1}_{d,l}^u$ is the relative path delay of the $l$-th path in the $d$-th cluster.
Moreover, $\theta_{d,l}^u$ and $\phi_{d,l}^u$ represent the AoD and the AoA corresponding to the $l$-th path of cluster $d$ of the channel. We assume the large-scale fading, AoAs and AoDs remain fixed for the $N_T$ time slots. Additionally, $\mathbf{a}_\text{TX}(.)$ and $\mathbf{a}_\text{RX}(.)$ are the array response vectors at the BS and UE $u$. We generate the specified parameters and functions as in \cite[Section IV-\textit{F}]{DL}. 

\begin{figure}
    \centering
    \begin{subfigure}[b]{0.45\textwidth}
         \centering
         \includegraphics[width=80mm]{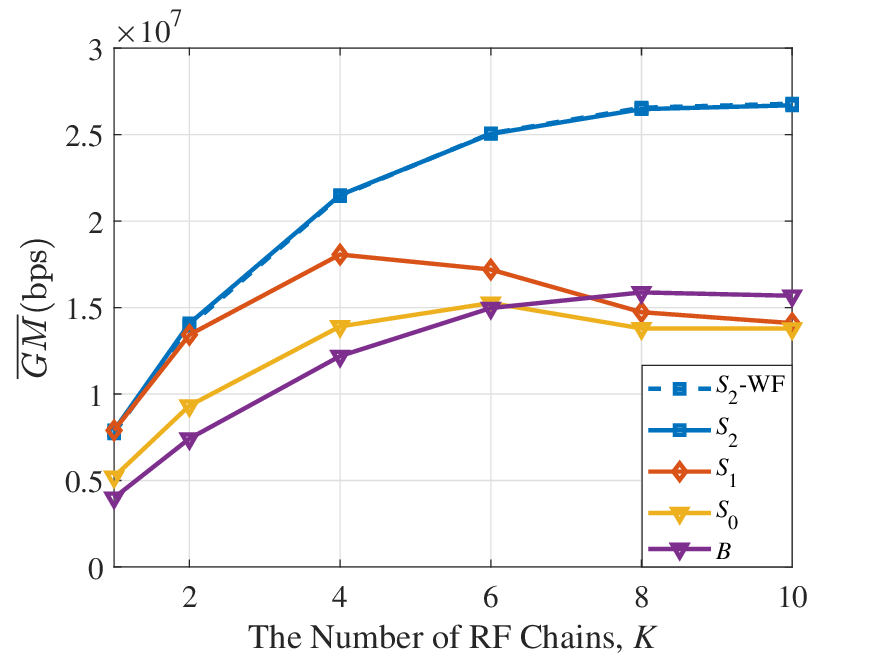}
         \caption{$U=10$}
         \label{r1.1}
    \end{subfigure}
    \hfill
    \begin{subfigure}[b]{0.45\textwidth}
         \centering
         \includegraphics[width=80mm]{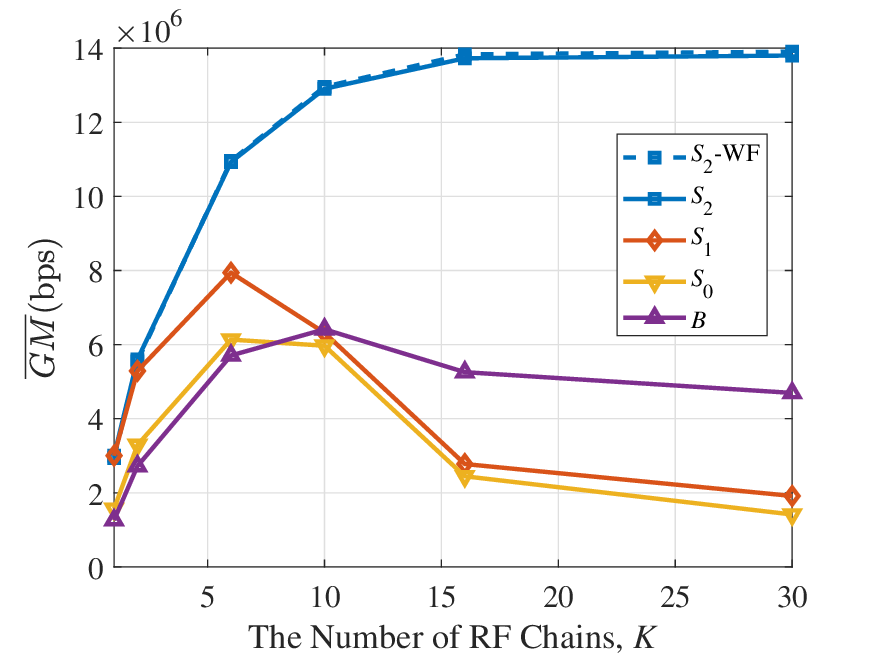}
         \caption{$U=30$}
         \label{r1.2}
    \end{subfigure}
    \caption{$\overline{GM}$ vs. $K$ with fixed $U$, $N_b=128$, $N_u=16$, $B_b=32$, $B_u=4$.}
    \label{r1}
    \vspace{-10pt}
\end{figure}

\begin{figure}
    \centering
    \begin{subfigure}[b]{0.45\textwidth}
         \centering
         \includegraphics[width=80mm]{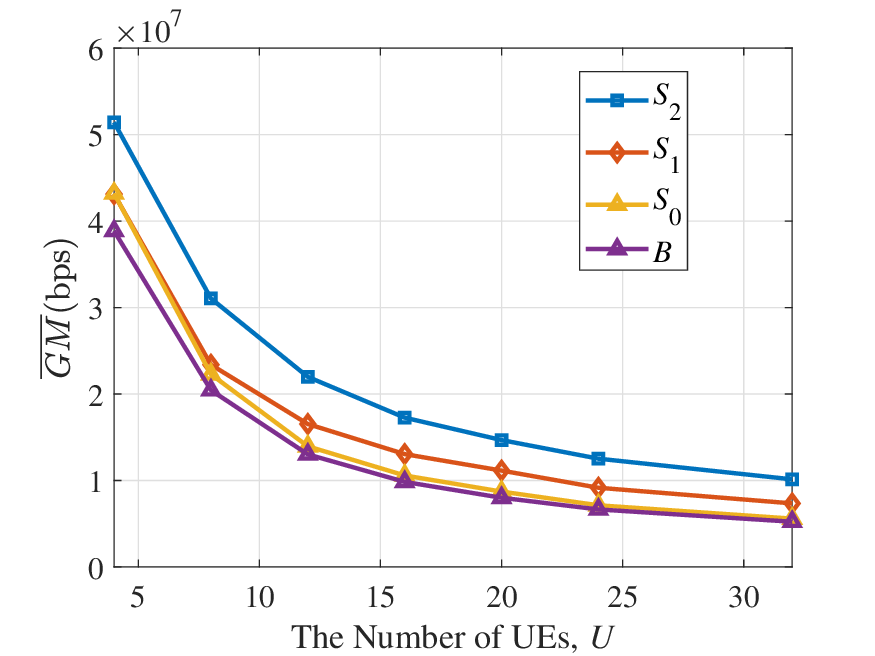}
         \caption{$K=6$}
         \label{r2.1}
    \end{subfigure}
    \hfill
    \begin{subfigure}[b]{0.45\textwidth}
         \centering
         \includegraphics[width=80mm]{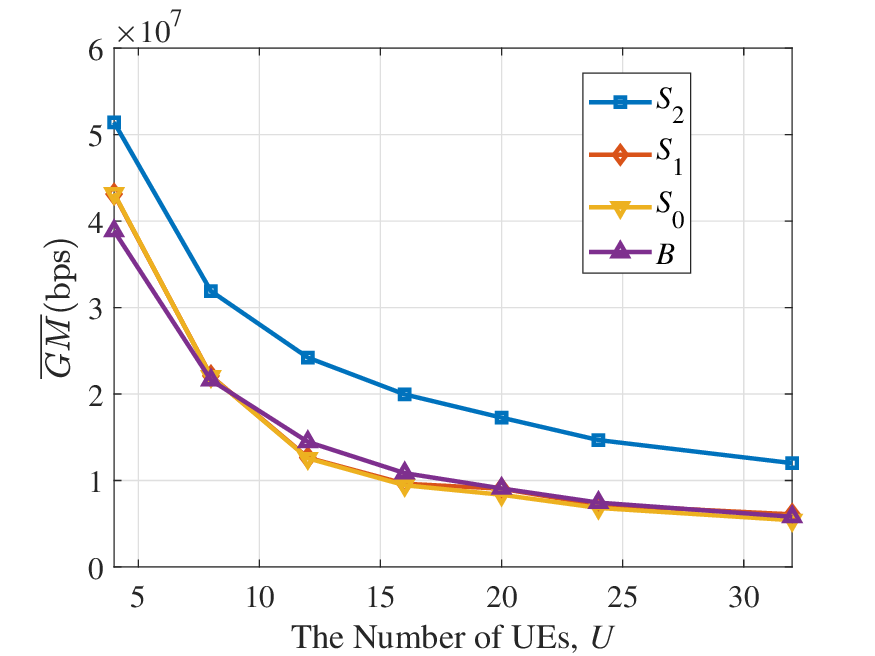}
         \caption{$K=10$}
         \label{r2.2}
    \end{subfigure}
    \caption{$\overline{GM}$ vs $U$ with fixed $K$, $N_b=128$, $N_u=16$, $B_b=32$, $B_u=4$.}
    \label{r2}

    \vspace{-10pt}
\end{figure}

We define a realization as $U$ uniformly distributed UEs with the corresponding $Q$ channel parameters per UE for each of $N_T=100$ time slots. The performance is the $GM$ of the UEs' rates over the $N_T=100$ time slots. In each realization, we perform BA, once at the beginning, and run the RRM solution for $N_T$ consecutive time slots. We set the initial $W$ as 10 and $R_u$ as 2 in the first time slot. After testing multiple values, %$X \in \{ 1, 2, 6, 12\}$ and $I_D \in \{\frac{1}{3}, \frac{2}{3}, 1, \frac{4}{3}, \frac{5}{3}, 2\}$, 
we found that $X=6=\frac{C}{Q}$ and an update step length of $X$ channels instead of one in Alg.~\ref{PerBeam} and $I_D=1$ in Alg.~\ref{IDD} can provide good performance and efficiency. Next, we provide the numerical results. $\overline{GM}$ is the $GM$ averaged over 50 different realizations in Fig~\ref{r1} and 200 different realizations in Fig~\ref{r2}.

% Remove EPA label

% \begin{figure}
%     \centering
%     \includegraphics[width=80mm]{Steps_10.eps}
%     \caption{The impact of different Blocks on the RRM solution with $U=10$, $N_b=128$, $N_u=16$, $B_b=32$, $B_u=4$.}
%     \label{r1}
% \end{figure}

% \begin{figure}
%     \centering
%     \includegraphics[width=80mm]{Steps_30.eps}
%     \caption{The impact of different Blocks on the RRM solution with $U=30$, $N_b=128$, $N_u=16$, $B_b=32$, $B_u=4$.}
%     \label{r2}
% \end{figure}

We first compare the performance of $B$, $S_0$, $S_1$ and $S_2$ for $U=10$   in Fig.~\ref{r1.1} and $U=30$ in Fig.~\ref{r1.2} where we plot $\overline{GM}$ as a function of $K$, the number of RF chains. We also show the performance of $S_2$ when we add the extra step with WF PA (the curve is labeled $S_2$-WF). PBS$^+$ (see $S_0$ versus $B$) increases the performance by 30\%  when $K=1$ and $U=10$. Note that PBS$^+$ does not do well when either $K$ is large or $U$ is large because PBS$^+$ schedules more users than PBS,  creating more interference and neither $B$ nor $S_0$ are IA. The LA BSel, WSRB,  (see $S_1$ versus $S_0$) improves the performance by 91\% wrt $S_0$ with $U=30$ and $K=1$ and by 29\% for $U=30$ and $K=6$. Note that $S_0$ and $S_1$ tend to select more UEs than $B$, which increases interference and this leads to a notable performance drop when $K$ is large. However, this side-effect is mitigated by the IA user dropping scheme (IDD), resulting in a non-decrease in performance when $K$ increases and an outstanding performance increase of 621\%  ($S_2$ versus $S_1$) for $U=30$ and $K=30$. Clearly $S_2$ is more useful when the number of users is larger. %The fact that the performance of $B$, $S_0$ and $S_1$ decreases as there are more RF chains also necessitates the importance of IDD when sharing a PRB between UEs. We can also find that IDD and LA BSel become more important as there are more UEs in the cell. 

Figs.~\ref{r1} also show that there is no need to add an extra WF PA step to $S_2$ since it does not impact performance. Finally, we note that  the performance of $S_2$ reaches a plateau when $K$ increases and that $K=6$ RF chains (resp. $K=10$) can achieve at least 90\% of the plateau when $U=10$ (resp. $U=30$) in $S_2$.  %We further validate our final heuristic $S_2$ by testing with fixed $K$ values of 6 and 10, assuming EPA, to assess performance across different numbers of UEs in the system.
Next, we show  the performance of our four solutions as a function of $U$, the number of UEs for two values of $K$, i.e., $K=6$ in Fig.~\ref{r2.1} and $K=10$ in Fig.~\ref{r2.2}.
In both figures, $S_2$ outperforms all the other solutions by roughly 38\% for $K=6$ and 97\% for $K=10$.  Therefore, neglecting interference limits the full potential of well-designed BSel and USel. %Only with IDD, which revisits the USel in an IA manner, the performance is always better with more RF chains.

Finally, we analyze the complexity of each step in our four heuristics in Table~\ref{CPLEX}, assuming we process in parallel PBS$^+$ for all the beams for $S_1$ and $S_2$ and PBS (resp. PBS$^+$) for $B$ (resp. $S_0$) for the $L$ beams. We also assume that we process in parallel   Alg.~\ref{IDD} for all the subchannels in a time slot. The main complexity is the USel since it adds channels to UEs in an iterative manner. WSRB BSel and IDD have negligible complexity because $|\mathcal{B}_p|<U \ll C U$. The overall complexity of all the four schemes are the same which confirms that $S_2$ is suitable for real-time operation.

\begin{table}
\centering
\setlength{\tabcolsep}{1.8mm}{
\caption{Complexity of the four solutions}
\label{CPLEX}
\footnotesize
\begin{tabular}{|c|c|c|c|c|} 

\hline

\rowcolor{gray} & USel & BSel & IDD & Overall\\
\hline

\cellcolor[HTML]{D3D3D3} $B$ & $\mathcal{O} (C U  )$ & $\mathcal{O} (L  )$ & $\times$ & $\mathcal{O} (C U  )$ \\
\hline

\cellcolor[HTML]{D3D3D3} $S_0$ & $\mathcal{O} (C U  )$ & $\mathcal{O}(L)$ & $\times$ & $\mathcal{O} (C U  )$ \\
\hline

\cellcolor[HTML]{D3D3D3} $S_1$ & $\mathcal{O} (C U  )$ & $\mathcal{O}(|\mathcal{B}_p|)$ & $\times$ & $\mathcal{O} (C U  )$ \\
\hline

\cellcolor[HTML]{D3D3D3} $S_2$ & $\mathcal{O} (C U   )$ & $\mathcal{O}(|\mathcal{B}_p|)$ & $\mathcal{O}(U)$ & $\mathcal{O} (C U  )$ \\
\hline

\end{tabular}}

\end{table}

\section{Conclusion} \label{Conclusion}

We have investigated the uplink radio resource management in a codebook-based HBF system considering multiple channels per time slot and a limited number of RF chains ($K<U$). We have shown that beam selection, user selection and power allocation are intricately coupled and that the sequence in which these steps are executed, to take part of the coupling(s) into account, significantly impacts the proportional fairness performance of the system. In particular, load-aware (LA) beam selection and interference-aware (IA) user selection have a huge impact on performance as shown by our extensive numerical campaign for a mmWave single cell.  Our LA and IA solution, $S_2$  performs six times better than the best of the other considered solutions and has a reasonable complexity. 

%A natural progression of this work is to consider digital beamforming for the RF chains which is left for future research. 

%Instead of solving a highly nonconvex optimization problem with inter-beam interference and the coupling between user selection and PA, we propose a low-complexity heuristic built on sequential blocks: per-beam user selection, load-aware (LA) beam selection, interference-aware (IA) user dropping and EPA. Our numerical results demonstrate the necessity of LA beam and user selections as purely independent and load-unaware beam selection and user selection (e.g. round robin) fail to provide reliable performance. We also found that IA user deselection can increase the performance by 89.4\% if there are multiple UEs in a subchannel, as it decreases the inter-beam interference not addressed in the per-beam user selection. Lastly, we show that EPA is as good as WF if beam and user selections are done properly.

\bibliographystyle{IEEEtran}
\bibliography{References.bib}

\end{document}